\documentclass[openacc]{rstransa}%%%%where rstrans is the template name

\titlehead{Research}

\usepackage{graphicx} % Required for inserting images

\usepackage{pifont}
\usepackage{color}
\usepackage{lineno}
\usepackage{tabularx}
\usepackage
{booktabs}

\usepackage
{courier}

\usepackage[version=4]{mhchem}
\newcommand\isotope[2]{\textsuperscript{#1}#2}

\begin{document}

%%%% Article title to be placed here

\title{Experimental neutrino physics in a nuclear landscape}

\author{%%%% Author details
D.~S.~Parno$^{1}$, A.~W.~P.~Poon$^{2}$ and V.~Singh$^{3}$}
%%%%%%%%% Insert author address here
\address{$^{1}$Department of Physics, Carnegie Mellon University, Pittsburgh, Pennsylvania 15213, USA\\
$^{2}$Nuclear Science Division, Lawrence Berkeley National Laboratory, Berkeley, California 94720, USA\\
$^{3}$Department of Physics, University of California, Berkeley, California 94720, USA}

%%%% Subject entries to be placed here %%%%
\subject{nuclear physics, neutrino physics}

%%%% Keyword entries to be placed here %%%%
\keywords{Neutrinos, neutrino mass, underground science, low radioactive background experimental techniques, neutrino mass ordering }

%%%% Insert corresponding author and its email address}
\corres{A.~W.~P.~Poon\\
\email{awpoon@lbl.gov}}

%%%% Abstract text to be placed here %%%%%%%%%%%%
\begin{abstract}

There are profound connections between neutrino physics and nuclear experiments. Exceptionally precise measurements of single and double beta-decay spectra illuminate the scale and nature of neutrino mass and may finally answer the question of whether neutrinos are their own antimatter counterparts. Neutrino-nucleus scattering underpins oscillation experiments and probes nuclear structure, neutrinos offer a rare vantage point into collapsing stars and nuclear fission reactors, and techniques pioneered in neutrino nuclear-physics experiments are advancing quantum-sensing technologies. In this article, we review current and planned efforts at the intersection of neutrino and nuclear experiments.
\end{abstract}
%%%%%%%%%%%%%%%%%%%%%%%%%%%

%%%%%%%%%% Insert the texts which can accommodate on firstpage in the tag "fmtext" %%%%%

\begin{fmtext}
\end{fmtext}

%\subsection{First subsection here}
%%%% Insert B head here
%Subsection text here.

%%%%%%%%%%%%%%% End of first page %%%%%%%%%%%%%%%%%%%%%

\maketitle
\section{Introduction}
\label{sec:intro}

Nuclear physics is pivotal in the story of the neutrino, the lightest known matter particle in the universe. Wolfgang Pauli first proposed the neutrino's existence in 1930 to solve the longstanding mystery of the nuclear beta-decay spectrum~\cite{Pauli:1930}, and Clyde Cowan and Frederick Reines discovered the particle experimentally at a nuclear reactor in 1956~\cite{Cowan:1956rrn}. At the same time, the neutrino is a tool for understanding nuclear physics, illuminating fusion reactions inside the Sun~\cite{Gann:2021ndb} and beta decays on Earth. Here, we consider nuclear physics to encompass phenomena dominated or driven by nuclear effects -- nuclear decays, nuclear reactions and nuclear structure -- especially at low energies.

In the past century of work, physicists have established~\cite{ParticleDataGroup:2022pth} that the neutrino is a neutral, left-handed lepton; the anti-neutrino, conversely, is right-handed. The three neutrino flavour states ($\nu_e, \nu_\mu, \nu_\tau)$, associated with charged-lepton flavours, are linear superpositions of the three neutrino mass states ($\nu_1, \nu_2, \nu_3$), with resultant flavour oscillation~\cite{Fukuda:1998mi, Ahmad:2002jz, Eguchi:2002dm} dictated by the mixing angles of the Pontecorvo–Maki–Nakagawa–Sakata (PMNS) matrix $U$, by the splittings $\Delta m_{ij}^2$ between mass states, and by the ratio of the source-detector distance to the neutrino energy. Since the neutrino's only Standard Model interaction is via the weak force, its interaction cross sections are intimidatingly small, but are measurable. Despite this progress, however, many basic questions are unanswered. For example, what is the absolute scale and nature (Dirac or Majorana) of the neutrino mass? What is the ordering of the mass values, and is there CP violation in the neutrino sector? Do the three known neutrino flavours constitute the whole neutrino sector, or are there ``sterile'' neutrinos that do not feel the weak force?

This review is conceived as a snapshot of current experimental efforts that relate nuclear and neutrino physics. We begin with an examination of two nuclear laboratories for exploring neutrino physics: single- and double-beta decay. As explored in Sec.~\ref{sec:numass}, nuclear beta decays (including electron-capture decays) permit a direct, kinematic probe of the absolute neutrino-mass scale. Recent work has significantly narrowed the laboratory limits on this quantity, a crucial input to both particle theory and cosmology. Meanwhile, in Sec.~\ref{sec:neutrinoless}, we see that searches for neutrinoless double beta decay -- a never-before-seen variant on the rare double beta-decay process, which is possible only if neutrinos are Majorana particles -- are taking nuclear-physics experiments to new scales and levels of background control. 
In Sec.~\ref{sec:othermeas}, we briefly survey additional intersections of neutrino and nuclear physics, 
including the nuclear physics of high-energy neutrino interactions, essential for interpreting long-baseline neutrino-oscillation experiments (Sec.~\ref{sec:othermeas}\ref{sec:high-energy-scattering}); the use of low-energy neutrino scattering to illuminate nuclear properties and supernova nucleosynthesis (Sec.~\ref{sec:othermeas}\ref{sec:low-energy-scattering}); neutrino probes of fission reactors (Sec.~\ref{sec:othermeas}\ref{sec:reactornu}); searches for sterile neutrinos (Sec.~\ref{sec:othermeas}\ref{sec:sterilenu}); and applications in quantum sensing (Sec.~\ref{sec:othermeas}\ref{sec:quantum_synergy}).

\section{Absolute neutrino-mass measurement}
\label{sec:numass}

Neutrino-oscillation experiments have established that neutrinos cannot all be massless, but are insensitive to the individual mass eigenvalues $m_i$. To date, oscillation data are consistent with two options for neutrino-mass ordering: $m_3 > m_2 > m_1$ (normal ordering) and $m_2 > m_1 > m_3$ (inverted ordering). In either ordering, the mass of the lightest neutrino sets a scale for the others, which can be probed via studies of $\beta$ and $\beta\beta$ decays (Sec.~\ref{sec:neutrinoless}), and via cosmological observations.  Some of these measurements are model-dependent. For example, the measured sum of the neutrino masses from cosmological studies --- $\sum_i m_i$ --- varies conspicuously under different model assumptions and data inputs, albeit achieving remarkable constraints~\cite{ParticleDataGroup:2022pth}. The measurements described in this section are essentially model-independent.    

Since Fermi established the kinematic relationship between the electron energy spectrum of the neutrino mass and $\beta$ decay~\cite{1934ZPhy...88..161F}, many experiments have tried to determine the neutrino mass via $\beta$ and electron-capture decays in different nuclei. Here, we focus on recent efforts and refer the reader to Ref.~\cite{Formaggio:2021nfz} for complete historical context.

Such probes of the neutrino-mass scale derive their sensitivity from precise spectral-shape measurements near the kinematic endpoint of the decay spectrum, where the presence of a non-zero neutrino mass appreciably changes the energy available to other particles in the final state.
In the quasi-degenerate regime, where the mass scale is large compared to the mass splittings, kinematic experiments measure an effective neutrino mass $m_\beta$:
\begin{equation}
    m^2_\beta = \sum_{i} \left| U_{e i} \right|^2 \, m_i^2 \; .
\end{equation}
Since the fraction of $\beta$-decays in the small, sensitive energy interval $\delta E$ below the endpoint energy $E_0 \approx Q$ (the $Q$-value of the decay), is proportional to $\left(\frac{\delta E}{Q} \right)^3$, an ideal nucleus for this type of measurement has a small $Q$-value, a relatively short half-life for the enhancement of source intensity, and a well-understood decay structure. 

The isotopes used in current neutrino-mass measurement efforts are \isotope{3}{H} (Sec.~\ref{sec:tritium}) and \isotope{163}{Ho} (Sec.~\ref{sec:holmium}). Each has some complexity in the measured spectrum.  In $\beta$-decay experiments with molecular \isotope{3}{H}, the final-state electronic, vibrational, and rotational excitations modify the beta spectrum significantly and are obtained from theory.   In \isotope{163}{Ho} electron-capture experiments, similarly intricate theoretical calculations are needed to account for X-ray, Auger-Meitner, and Coster-Kronig transitions, as well as nuclear recoil.  

Recent efforts have identified other possible isotopes with ultralow $Q$-values ($<1$~keV), and thus enhanced statistical sensitivity; see Ref.~\cite{Keblbeck:2022twm} for a review.
These isotopes, however, typically have extremely long lifetimes, complex nuclear structures, and very small branching ratios for the specific decay modes with ultralow $Q$, rendering them impractical targets for future precise experiments. 

\subsection{\isotope{3}{H}}
\label{sec:tritium}
The best current kinematic limit on the neutrino-mass scale arises from the decay of molecular tritium, \ce{{\isotope{3}{H}}2}:
\begin{equation}
    \ce{{\isotope{3}{H}}2} \, \longrightarrow \, \ce{{\isotope{3}{He}\,\isotope{3}{H}}}^+ + \text{e}^- + \bar{\nu}_\text{e} \, + \, Q(\ce{{\isotope{3}{H}}2}) \; .
\end{equation}
The differential decay rate, summed over all final molecular states $f$ in the daughter molecule, each with energy $V_f$ and weighted by the transitional probability $P_f$ to that state, is~\cite{Kleesiek:2018mel}:
\begin{align}
        \nonumber
    \frac{\text{d} \Gamma}{\text{d} E} &= \frac{G_\text{F}^2 \, |V_\text{ud}|^2}{2\pi^3} \; |M_\text{nuc}|^2 \; F(Z, E) \cdot p\, (E+m_e) \\
      &\quad \cdot \sum_f \; P_f \; \epsilon_f \; \sqrt{ \epsilon_f^2 - m^2_\beta } \; \Theta(\epsilon_f - m_\beta) \; ,
    \label{eq:diffspec}
\end{align}
where $G_\text{F}$ is the Fermi coupling constant, and $|V_\text{ud}|=0.97373 \pm 0.00031$ is the CKM matrix element~\cite{ParticleDataGroup:2022pth}.  $M_\text{nuc}$ is the nuclear transition matrix element. $F(Z, E)$ is the Fermi function that accounts for the Coulomb interaction between the outgoing electron with kinetic energy $E$ and momentum $p$, and the daughter nucleus with atomic charge $Z$; $\epsilon_f$ is the neutrino energy $(= E_0 - V_f - E)$; $E_0$ is the maximum $\beta$ energy if the neutrino mass is zero, and the Heaviside step function $\Theta(\epsilon_f - m_\beta)$ ensures energy conservation.  

\begin{figure}
\begin{center}
\begin{tabular}{cc}
\includegraphics[width=0.45\textwidth]{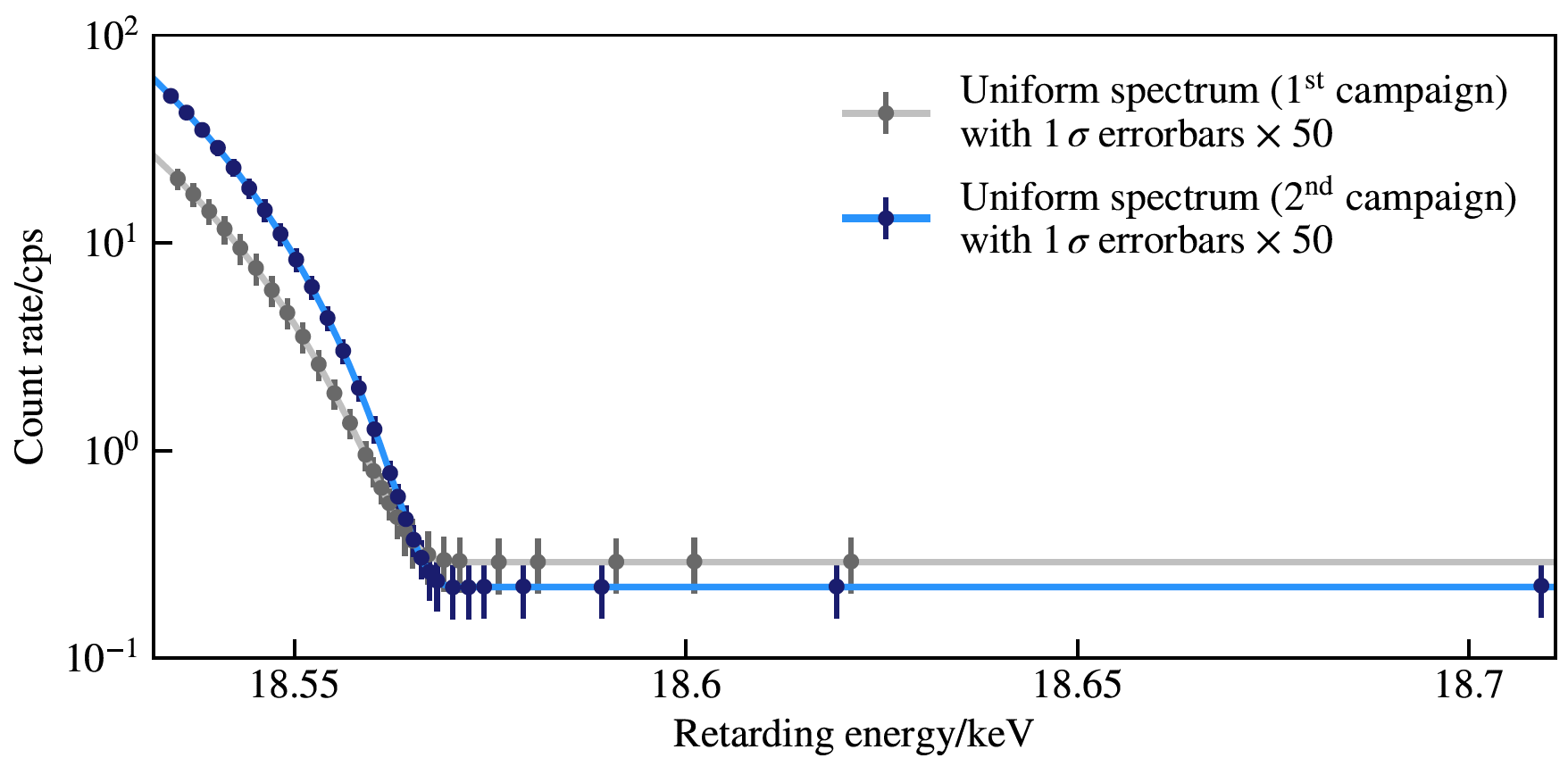} &
\includegraphics[width=0.45\textwidth]{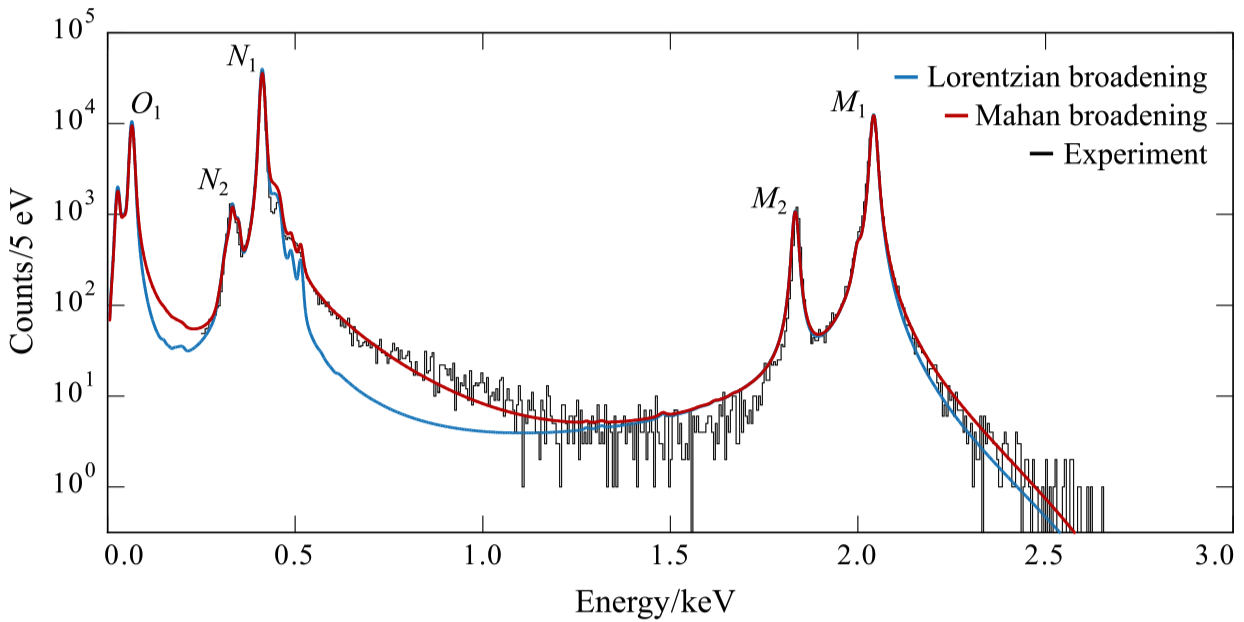} 
\end{tabular}
\end{center}
\caption{Measured integral \isotope{3}{H} $\beta$ spectrum near the endpoint from the KATRIN experiment~\cite{KATRIN:2021uub} (left) and the differential electron-capture calorimetric spectrum of \isotope{163}{Ho} from the ECHo experiment~\cite{Velte:2019jvx} (right).\label{fig:spectra-compare}}
\end{figure}

The KArlsruhe TRItium Neutrino (KATRIN) experiment~\cite{KATRIN:2021dfa} is the most sensitive operating direct neutrino-mass experiment, and the current best limit of $m_\beta < 0.8$~eV/c$^2$ (90\% C.L.) is based on its first two measurement campaigns~\cite{KATRIN:2021uub}.   
In the KATRIN apparatus, cold \ce{{\isotope{3}{H}}2} gas is injected into the windowless source section.  The decay electrons are guided by magnetic fields to the main spectrometer for energy analysis. Differential and cryogenic pumping stages along the beamline reduce the tritium flow by 14 orders of magnitude.  The electrons' transverse momentum is adiabatically transformed into longitudinal momentum in a slowly varying magnetic field, which reaches a minimum in the ``analysing'' plane of the main spectrometer. Only electrons with enough kinetic energy to pass the potential barrier of ~$\sim -18.6$~kV are transmitted to the detector. Essentially, the main spectrometer acts as a high-pass filter so that the detector records an integral spectrum (Fig.~\ref{fig:spectra-compare}, left). KATRIN continues to take data toward a design sensitivity goal of $m_\beta < 300$~meV.

The statistical sensitivity of a KATRIN-type experiment follows~\cite{Otten:2008zz}:
\begin{equation}
\delta m^2_\beta \propto \frac{b^\frac{1}{6}}{r^\frac{2}{3} t^\frac{1}{2}}
\end{equation}
where $b$ is the background rate, $r$ is the radius of the spectrometer, and $t$ is the measurement time. KATRIN observes higher-than-expected backgrounds arising from low-energy electrons generated in the large volume of the main spectrometer; even without this issue, the diameter of a spectrometer to improve on KATRIN's sensitivity by an order of magnitude would be unrealistically large. Instead of enlarging the spectrometer, improvements to a KATRIN-like experiment would thus require reducing the background or changing the role of the main spectrometer in the measurement, e.g. by operating in a time-of-flight mode~\cite{KATRIN:2022ayy}.

It is clear from Eq.~\ref{eq:diffspec} that the molecular final-state distribution (FSD) populated by \ce{{\isotope{3}{H}}2}
decay affects the measured $\beta$ spectrum, and hence $m^2_\beta$.  Near the spectral endpoint, any unaccounted-for Gaussian broadening  $\Delta\sigma$, including that of the FSD, changes the extracted $m^2_\beta$ by~\cite{Robertson:1988xca}:
\begin{equation}
\label{eq:unaccountedvariance}
\Delta m^2_\beta \approx -2 \Delta \sigma^2.
\end{equation}
In fact, it has been demonstrated using Eq.~\ref{eq:unaccountedvariance} that the use of a more sophisticated, modern FSD~\cite{Saenz:2000dul} is enough to render negative $m^2_\beta$ results from the 1980s consistent with zero~\cite{Bodine:2015sma}. Significant theoretical effort has been invested in improved FSD calculations, consistent with the limited available experimental tests~\cite{Bodine:2015sma, TRIMS:2020nsv}, and the resulting uncertainty is now negligible for KATRIN. However, the spectral broadening induced by the FSD would become a significant limiting factor for experiments with sufficiently sharp energy resolution.

Cyclotron Radiation Emission Spectroscopy (CRES), an alternative to the KATRIN strategy, is a frequency technique for determining $m_\beta$ by 
precisely measuring the cyclotron radiation from the relativistic electron in atomic \isotope{3}{H} $\beta$ decay~\cite{Monreal:2009za}. The power radiated by an 18-keV electron in a 1~T field is approximately 1~fW.  The Project~8 experiment aims to realise this concept by measuring the differential energy spectrum of atomic \isotope{3}{H}, thereby eliminating the FSD broadening of molecular \isotope{3}{H}$_2$ and the need for a mammoth spectrometer.  

In its first milestone, the Project~8 experiment observed single electrons from \isotope{83m}{Kr} decay~\cite{Project8:2014ivu}.  Recently, the experiment has demonstrated CRES as a viable technique for a low-background neutrino-mass measurement with \isotope{3}{H}$_2$ in a small trap, setting a Bayesian upper limit of $m_\beta <$155~eV/$c^2$ (90\% C.L.).  No background was observed after 82~days of running, and an adequate resolution was demonstrated using \isotope{83m}{Kr} 17.8-keV internal-conversion electrons~\cite{Project8:2022hun, Project8:2023jkj}. The collaboration is now following two parallel research-and-development (R\&D) tracks: scaling up 
 the CRES technique to larger volumes with resonant cavities, and developing an atomic tritium source in which magnetic trapping of \isotope{3}{H} atoms prevents recombination. 
  The ultimate goal of the Project 8 experimental programme is a sensitivity of $m_\beta < $~40~meV/c$^2$ (90\% C.L.) with atomic \isotope{3}{H}.  

\subsection{\isotope{163}{Ho}}
\label{sec:holmium}

 Serious non-tritium-based efforts to the neutrino mass currently centre on 
 the internal-bremsstrahlung electron-capture (IBEC) decay spectrum of~\isotope{163}{Ho} to~\isotope{163}{Dy}, as first proposed in Ref.~\cite{DeRujula:1982qt}. Earlier calorimetric measurements of the \isotope{187}{Re} beta-decay spectrum have been abandoned due to difficulties in designing a suitable, scalable detector~\cite{Nucciotti:2015rsl}. Despite its relatively low Q-value, \isotope{187}{Re} has an exceptionally long half-life ($4.3\times10^{10}$~y) due to its first-order forbidden transition. But the \isotope{187}{Re} programme had significantly advanced the development of the  \isotope{163}{Ho} experiments; in particular, in microcalorimetry.

In~\isotope{163}{Ho} measurements, microcalorimeters capture the de-excitation electrons and photons of the daughter \isotope{163}{Dy}* entirely and convert them to heat.  The neutrino mass is manifested in the upper end of the ~\isotope{163}{Dy}* deexcitation spectrum, similar to the modification in $\beta$-decay spectra near the endpoint.  The synthetic isotope \isotope{163}{Ho} has a low $Q=2.833$~keV~\cite{ECHo:2015qgh} and a half-life of 4570~years.  Since neutrinos are emitted in electron-capture decays instead of antineutrino emission in \isotope{3}{H} decay, the two types of measurement are complementary. The \isotope{163}{Ho} spectrum has a complex shape (Fig.~\ref{fig:spectra-compare}) requiring a careful treatment of resonance features, shake-off electrons, and solid-state effects from the absorber~\cite{Brab:2020uzx}. The pile-up rate in the microcalorimeters can be very high as they measure the full differential spectrum.  Different  \isotope{163}{Ho} experiments deploy different SQUID-based detector technologies to read out minute temperature changes.   HOLMES~\cite{HOLMES:2016spk} uses Transition Edge Sensor (TES) arrays, while ECHo uses arrays of magnetic metallic calorimeters (MMCs) and has set a limit of $m_\beta < 150$ eV/c$^2$ (95\% C.L.)~\cite{Velte:2019jvx}. HOLMES has recently measured its first \isotope{163}{Ho} decay spectrum. The ECHo-1k phase of the programme will reach a neutrino-mass sensitivity below 20 eV/c$^2$; its next phase, ECHo-100k,is projected to reach a sensitivity below 2~eV/c$^2$~\cite{Griedel:2022xzj}. An international consortium has recently called for an effort of neutrino-mass measurement in \isotope{163}{Ho} with sub-eV sensitivity~\cite {Ullom:2022kai}.

\section{Neutrinoless double-beta decay}
\label{sec:neutrinoless}

It is not currently understood how the existence of neutrino mass should be incorporated into the Standard Model. Since the neutrino has no electric charge, both Dirac and Majorana mass terms are possible. The search for neutrinoless double-beta decay ($0\nu\beta\beta$) is the only practical means to establish which best describes the neutrino nature. 

In the Standard Model, double-beta decay ($2\nu\beta\beta$) is an allowed second-order decay process in which two uncorrelated nucleons decay simultaneously and emit two electrons and two anti-neutrinos: 
$(Z,A) \xrightarrow[]{} (Z+2,A) + 2e^- + 2\Bar{\nu}$.
However, if neutrinos are their own antiparticles (Majorana particles~\cite{majoranasymmetric}), the emitted anti-neutrino from one of the nucleons can be absorbed in the second interaction so that there are no neutrinos in the final state:
\begin{equation}
\label{eqn:0nubb}
(Z,A) \xrightarrow[]{} (Z+2,A) + 2e^- .
\end{equation}

\begin{figure}[htbp]
\begin{center}
\includegraphics[width=0.9\textwidth]{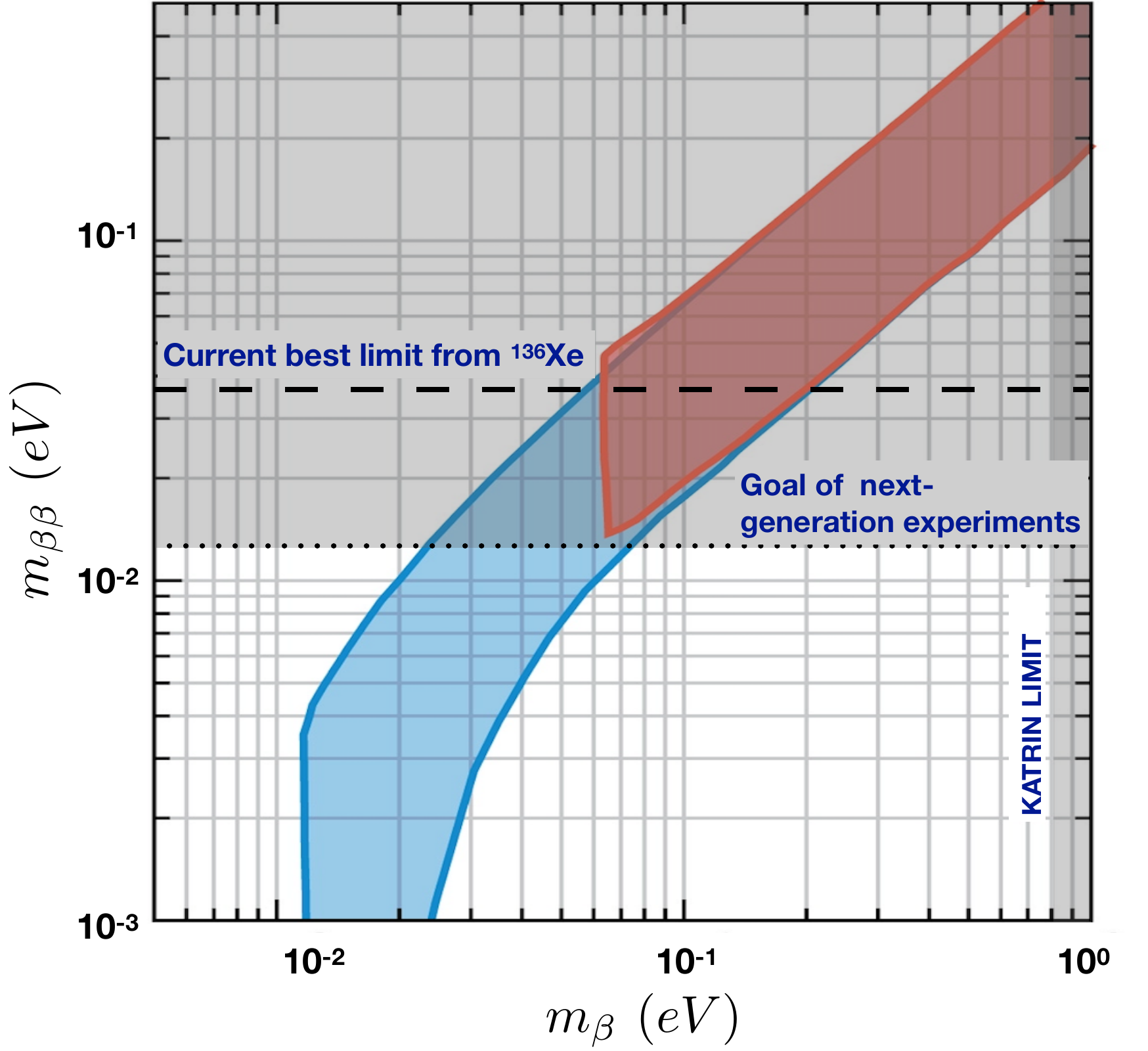}
\caption{The effective Majorana-mass observable $m_{\beta\beta}$ in neutrinoless double-beta decay searches vs. the direct kinematic observable $m_\beta$.  The neutrino mixing parameters $U_{\alpha i}$  are varied within their ranges from oscillation experiments. The blue area is for the normal mass ordering, while the red area is for the inverted mass ordering. The next generation of $0\nu\beta\beta$ experiments aims to probe the entire inverted mass ordering through $m_{\beta\beta}$. Adapted from~\cite{Dolinski:2019nrj}.}
\label{fig:mbb_vs_mb_mod1}
\end{center}
\end{figure}

This is the much sought-after $0\nu\beta\beta$ decay mode, which violates lepton-number conservation -- an accidental symmetry of the Standard Model -- and could help explain the matter-antimatter asymmetry in our Universe~\cite{Fukugita:1986hr}. The $0\nu\beta\beta$ decay rate can be expressed as
\begin{equation}
\label{eqn:0nubbDecayRate}
[T_{\frac{1}{2}}^{0\nu}]^{-1} = \sum_{i} G_{i}^{0\nu}(Z,Q) \cdot \left|M_{i}^{0\nu}\right|^{2} \cdot \zeta_{i}^2
\end{equation}
where $G^{0\nu}(Z,Q)$ is the phase-space factor that depends on the proton number ($Z$) of the decaying nucleus and the $Q$-value of the decay, $M_{i}^{0\nu}$ is the nuclear matrix element (NME), and $\zeta_{i}$ depends on the mechanism and mode of the lepton-number-violating process. The phase-space factors have been calculated~\cite{Kotila:2012zza, Deppisch:2020ztt} and the $Q$-values have been measured precisely for several isotopes actively pursued by experiments~\cite{Mount:2010zz,Rahaman:2007ng,Fink:2011yx,Scielzo:2009nh}. If we assume that the decay is mediated by the exchange of light Majorana neutrinos, $\zeta$ reduces to an effective Majorana neutrino mass $m_{\beta\beta}$, which is a coherent sum of neutrino mass eigenvalues defined as
\begin{equation}
\label{eqn:mbb}
|m_{\beta\beta}|=|\sum_{i=1}^3 U^2_{ei} m_i|
\end{equation}
Fig.~\ref{fig:mbb_vs_mb_mod1} shows the relationship between $m_{\beta\beta}$ and $m_{\beta}$ (Sec.~\ref{sec:numass}). 

In the scenario of light-neutrino exchange, the decay rate is written as~\cite{Engel:2016xgb,Simkovic:2021tkr}:
\begin{equation}
\label{eqn:0nubbDecayRate_light}
[T_{\frac{1}{2}}^{0\nu}]^{-1} =G^{0\nu}(Z,Q) \cdot (g_{A})^4 \cdot \left|M^{0\nu}\right|^{2} \cdot \frac{m_{\beta\beta}^2}{m_{e}^2}
\end{equation}
where $g_{A}$ is the axial-vector coupling constant factored out of the nuclear matrix element $|M^{0\nu}|^{2}$, and $m_e$ is the mass of the electron. The corresponding NMEs are calculated using various macroscopic and microscopic nuclear models dealing with complex nuclear structures (Sec.~\ref{sec:nme}). However, the predictions from these nuclear models disagree by more than a factor of two~\cite{Engel:2016xgb}, which results in a significant uncertainty on the predicted value of $m_{\beta\beta}$. 
While the constraint on neutrino mass through $0\nu\beta\beta$ is model-dependent, establishing the Majorana character is not; the Black-Box Theorem~\cite{Schechter:1981bd,Nieves:1984sn,Takasugi:1984xr} states that the observation of neutrinoless double-beta decay would directly imply lepton-number violation.

\begin{figure}[htp]
    \centering
    \includegraphics[width=\textwidth]{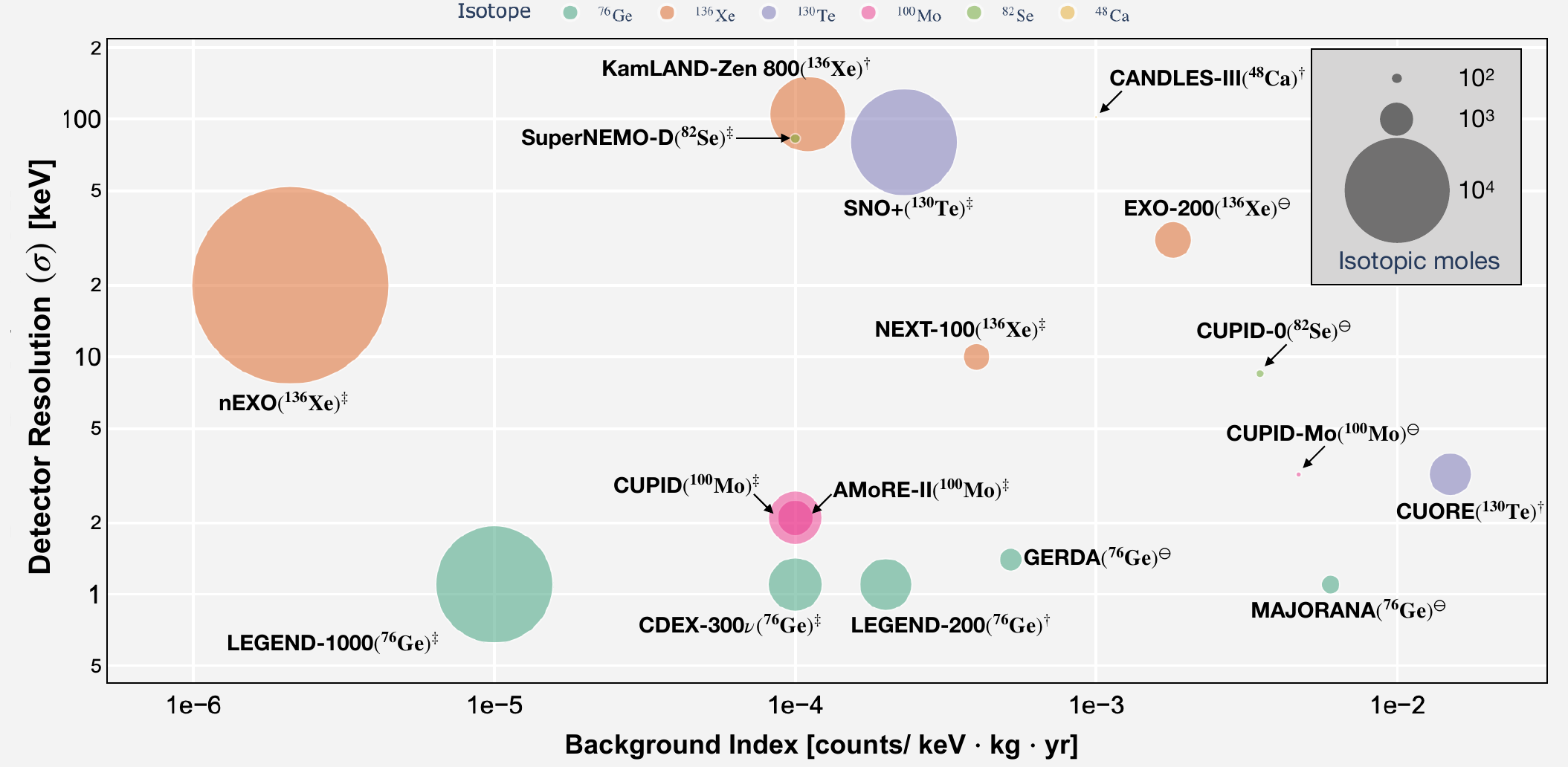}
    \caption{Relevant experimental parameters -- background index, detector resolution ($\sigma$), and isotopic moles -- for recently completed ($\ominus$), currently running ($\dag$), and proposed ($\ddag$) $0\nu\beta\beta$-decay search experiments. Furthermore, isotopes with high $Q_{\beta\beta}$ value offer an additional advantage since the phase space factor $G^{0\nu}(Z,Q)$ is proportional to $Q_{\beta\beta}^5$ and radioactive backgrounds tend to be smaller at higher energy.}
    \label{fig:res_bkg_exp}
\end{figure}

The fundamental concept behind a $0\nu\beta\beta$ search involves detecting two emitted electrons and identifying their summed energy peak at the $Q$-value ($Q_{\beta\beta}$) of the energy spectrum. Typically, the $Q$-values of relevant isotopes are precisely known
and the search for a $0\nu\beta\beta$ peak is limited to a narrow energy range determined by the detector's energy resolution ($\Delta E$) at $Q_{\beta\beta}$. The search sensitivity is also limited by background events that mimic the signal signature in the region of interest (ROI). The half-life sensitivity of an experiment can be expressed as~\cite{Moe:1991ku}:
\begin{align}
\label{eqn:0nusensitivity}
T_{1/2}^{0\nu} & \propto \epsilon \cdot \sqrt{\frac{M \cdot t}{B \cdot {\Delta E}}} & \text{background-limited} \nonumber \\
T_{1/2}^{0\nu} & \propto \epsilon \cdot {M \cdot t}&\text{background-free} 
\end{align}
where $\epsilon$ is the detector's efficiency, $M$ is the mass of the isotope deployed, and $B$ is the background index, typically expressed as the number of background events expected in a certain energy range within the live-time of the experiment ($t$) for a given detector mass. It is often reported in units of counts per detector mass, energy, and time, \textit{e.g.}\, counts/(keV$\cdot$kg$\cdot$year).

Given the extreme rarity of this decay, the experimental challenge lies in detecting this process amidst a background of other radioactive decays and cosmic rays~\cite{Formaggio:2004ge}. Experiments are carried out in deep underground facilities~\cite{Ianni:2017vqi} that provide a natural barrier against cosmic-ray interference. The dominant sources of radioactive background typically include $\alpha$, $\beta$, and $\gamma$ radiation from primordial decay chains, together with neutron-induced reaction products in underground labs. Experiments using isotopes with $Q_{\beta\beta}$ > 2615 keV benefit from a lower background by avoiding the $^{208}$Tl line. 
 However, it is not always possible to deploy these isotopes on a large scale due to their low isotopic abundance or a lack of suitable detector technology with low enough background levels. 
 
 Some of the world's leading limits on $0\nu\beta\beta$ decay were obtained with experiments using $^{76}$Ge, $^{136}$Xe, and $^{130}$Te -- all with  $Q_{\beta\beta}$ < 2615~keV -- using very powerful detection and background-rejection techniques.  These experiments use materials with low radioactive content in detector construction, minimising internal background sources while employing layers of passive shielding to reduce external backgrounds. 
Experiments employing water or liquid cryogens for passive shielding can use the same medium as an active veto for cosmic rays. Active background rejection, using such techniques as timing, event topology, fiducialisation of the active detector volume, and particle identification, complements passive methods.

 \begin{table}[!h]
 \centering
 \caption{ Comparison of lower half-life limits $T_{1/2}^{0\nu}$ (90\% CL) and corresponding $m_{\beta\beta}$ limits for the recently completed, currently running, and next-generation proposed experiments. Each range of $m_{\beta\beta}$ upper limits is as reported by that experiment and depends on their choice of multiple matrix elements. The measured sensitivities are reported in bold for contrast with projected sensitivities.}%%%Table caption goes here
\label{tab:ndbd_experiments}
\begin{tabularx}{\textwidth}{p{3cm}X p{1.8cm}X X X X }
\rowcolor{gray}
Experiment & Status & {Isotope} & {$T_{1/2}^{0\nu}$ [yr]}  & {$m_{\beta\beta}$ [meV]} 
\\\hline
\addlinespace[0.5ex]
{GERDA}~\cite{GERDA:2020xhi} & Completed & $^{76}\text{Ge}$ & $\mathbf{1.8 \times 10^{26}}$ & \textbf{79---180} \\
 {MAJORANA}~\cite{Majorana:2022udl} & Completed &$^{76}\text{Ge}$ & $\mathbf{8.5 \times 10^{25}}$ & \textbf{113---269} \\
 {LEGEND-200}~\cite{LEGEND:2021bnm} & Taking Data & $^{76}\text{Ge}$ & $1.5 \times 10^{27}$ & 34---78 \\
 {LEGEND-1000}~\cite{LEGEND:2021bnm} & Proposed & $^{76}\text{Ge}$ & ${8.5 \times 10^{28}}$ & 9---21 \\
 {CDEX-300$\nu$}~\cite{CDEX:2023owy} &  Proposed & $^{76}\text{Ge}$ & $3.3 \times 10^{27}$ & 18---43 \\
 {KamLAND-Zen}~\cite{KamLAND-Zen:2022tow} & Taking Data & $^{136}\text{Xe}$ & $\mathbf{2.3 \times 10^{26}}$ & \textbf{36---156} \\
 {EXO-200}~\cite{EXO-200:2019rkq} & Completed &$^{136}\text{Xe}$ & $\mathbf{3.5 \times 10^{25}}$ & \textbf{93---286} \\
 {nEXO}~\cite{nEXO:2021ujk} & Proposed & $^{136}\text{Xe}$ & $1.3 \times 10^{28}$ & 6.1---27 \\
 {NEXT-100} & Construction & $^{136}\text{Xe}$ & $7.0 \times 10^{25}$ & 66---281 \\
 {CUORE}~\cite{CUORE:2021mvw} & Taking Data &$^{130}\text{Te}$ & $\mathbf{2.2 \times 10^{25}}$ & \textbf{90---305} \\
 {SNO+}~\cite{Chen:2021llf} & Construction & $^{130}\text{Te}$ & $2.1 \times 10^{26}$ & 37---89 \\
 {AMoRE-II}~\cite{Kim:2022uce} & Proposed & $^{100}\text{Mo}$ & $5.0 \times 10^{26}$ & 17---29 \\
 {CUPID-Mo}~\cite{Augier:2022znx} & Completed & $^{100}\text{Mo}$ & $\mathbf{1.8\times 10^{24}}$ & \textbf{280---490} \\
 {CUPID}~\cite{CUPID:2019imh} & Proposed & $^{100}\text{Mo}$ & $1.5\times 10^{27}$ & 10---17 \\
 {CUPID-0}~\cite{CUPID:2022puj} & Completed & $^{82}\text{Se}$ & $\mathbf{4.6\times 10^{24}}$ & \textbf{263---545} \\
 {SuperNEMO-D}~\cite{Quinn:2023rcl} & Construction & $^{82}\text{Se}$ & $4.0\times 10^{24}$ & 260---500 \\
 {CANDLES-III}~\cite{CANDLES:2020iya} & Taking data & $^{48}\text{Ca}$ & $\mathbf{5.6\times 10^{22}}$ & \textbf{2900---1600} \\

\hline

\end{tabularx}
\end{table}

Figure~\ref{fig:res_bkg_exp} shows the recently completed, running, and proposed experiments in terms of achieved or projected background indices, detector resolution, and the amount of target isotope used. A comparison of the lower half-life limits and corresponding $m_{\beta\beta}$ values is given in Table~\ref{tab:ndbd_experiments}. Although $0\nu\beta\beta$ decay has not been observed, the current generation of experiments has successfully deployed several hundreds of kilogrammes of isotopes to push the $T_{1/2}^{0\nu}$ lower limit in the order of $10^{25}$---$10^{26}$ years despite being background-limited. The next generation of experiments seeks to increase the isotope mass significantly ($\approx$ tonne-scale). Since the sensitivity in a background-free experiment scales linearly with measurement time $t$ instead of $\sqrt{t}$, these experiments aim to reduce background levels by several orders of magnitude. Deploying a ``ton-scale'' detector and considerably decreasing the background will allow next-generation experiments to probe half-lives in the range of $10^{26}$---$10^{28}$~years with typically ten years of data taking, setting a limit of $m_{\beta\beta} < (18.4 \pm 1.3) $~meV~\cite{Agostini:2021kba} and ruling out the inverted hierarchy of neutrino mass spectra. 
We reiterate that this interpretation of half-life limits in terms of $m_{\beta\beta}$ is only valid under the standard assumption that the decay is mediated by light Majorana neutrino exchange, and all other mechanisms that could result in $0\nu\beta\beta$ decay are considered negligible or nonexistent.

It is essential to search for $0\nu\beta\beta$ in multiple isotopes not only for confirmation of discovery but also to identify the underlying mechanism of decay~\cite{Deppisch:2006hb,Bossio:2023wpj, Lisi:2023amm,Graf:2022lhj}. Experiments using different isotopes provide varying search sensitivity and background discrimination capabilities and are subject to independent systematic uncertainties. Measurements of multiple isotopes, therefore, not only allow cross-validation of theoretical models across different nuclear environments but also help identify and control experimental systematic uncertainties. In addition, precise measurements of diverse two-neutrino processes are needed to refine NME calculations and provide a comprehensive grasp of the Standard Model background in $0\nu\beta\beta$ searches. Consequently, technologies that can reconstruct individual energies and topologies (e.g., SuperNEMO~\cite{Quinn:2023rcl}) will have a crucial role even if they may not allow significant scaling up of isotopic mass. Analogously, the light $\beta\beta$ emitter $^{48}$Ca, which has extremely low isotopic abundance (0.2\%) and will be extremely cost-intensive to scale up, is an interesting nucleus to probe since it is an ideal target for benchmarking various NME calculations (Sec.~\ref{sec:nme}). 

Numerous review articles summarise the massive experimental and theoretical efforts over the last couple of decades~\cite{Agostini:2022zub, Dolinski:2019nrj, DellOro:2016tmg, Vergados:2012xy, Avignone:2007fu, Elliott:2002xe}. Here, we take a bird's-eye view of the experimental landscape for $0\nu\beta\beta$-decay searches, focusing on efforts toward future ton-scale experiments. We refer the reader to Fig.~\ref{fig:res_bkg_exp} and Table~\ref{tab:ndbd_experiments} for the relevant experimental parameters. Different isotopes offer distinct advantages and challenges regarding experimental criteria of scalable detector technology that can provide comprehensive sensitivity and background discrimination.

\subsection{\texorpdfstring{$^{76}$Ge ($Q_{\beta\beta}$ = 2039.0 keV)}{}}

The GERmanium Detector Array (GERDA) experiment~\cite{GERDA:2020xhi} was located at the Laboratori Nazionali del Gran Sasso (LNGS), Italy. Its high-purity germanium detectors (HPGe) 
were enriched to 87\% of the isotope $^{76}$Ge and immersed and cooled directly in ultra-pure liquid argon (LAr), whose scintillation also actively vetoed radioactive backgrounds.
GERDA achieved the lowest background sensitivity (BI$\times \sigma$) obtained by any $0\nu\beta\beta$-decay search. 
The {\sc Majorana Demonstrator} (MJD)~\cite{Majorana:2022udl} at the Sanford Underground Research Facility (SURF), USA, took a more conventional approach with layered ultralow-background electroformed copper (EFCu) and lead shielding for HPGe detectors housed in EFCu vacuum cryostats. 
MJD reported the best detector resolution among all $0\nu\beta\beta$-decay searches with $\sim$1.1~keV ($\sigma$) at $Q_{\beta\beta}$.  

The next-generation Large Enriched Germanium Experiment for Neutrinoless $\beta\beta$ Decay  (LEGEND)~\cite{LEGEND:2021bnm} aims to achieve a sensitivity of $T_{\frac{1}{2}}^{0\nu} > 10^{28}$~yr by combining the best of GERDA and MJD technologies. The goal is to operate 1 ton of enriched germanium detectors for 10 years at a background index of $\sim1\times10^{-5}$ counts/kg$\cdot$keV$\cdot$yr. The programme is being pursued in phases, with LEGEND-200 currently deployed and taking data in an upgraded GERDA infrastructure. 
%LEGEND-200 will use 70~kg detectors from GERDA and MJD experiments, while the other 130~kg of mass has been freshly procured. 
LEGEND-200's projected background index is about three times lower than GERDA, mainly due to fewer cables and electronic components per unit mass of HPGe detectors, an improved light readout for the liquid-argon veto, and improvements in the radiopurity of construction materials. LEGEND-1000 envisages reducing the background index by another factor of 20 by using underground argon (with a reduced level of radioactive \isotope{42}{Ar}) for shielding, and further reducing the radioactivity levels in components in the vicinity of the detectors. 

A similar large-scale Ge effort is being pursued for the CDEX-300$\nu$ experiment~\cite{CDEX:2023owy}. Like GERDA and LEGEND, the detectors will be immersed in liquid argon that serves as both a cooling medium and veto detector. CDEX-300$\nu$ is expected to demonstrate a half-life sensitivity of $T_{\frac{1}{2}}^{0\nu} > 3.3\times10^{27}$~yr with an effective runtime of 10 years. Small prototypes with a detector mass of $\sim$1~kg have been deployed. 

\subsection{\texorpdfstring{$^{136}$Xe ($Q_{\beta\beta}$ = 2457.8 keV)}{}}

The KamLAND-Zen~\cite{KamLAND-Zen:2022tow} (Kamioka Liquid Scintillator Anti-Neutrino Detector Zero-Neutrino) series of experiments is located at the Kamioka Observatory, Japan. Its KamLAND-Zen~800 phase uses 745~kg of Xe gas, enriched to 90-91\% and dissolved at 3\% by weight into liquid scintillator in a nylon inner balloon, which in turn is surrounded by 1~kilo-ton of liquid scintillator acting as an active shield.  $^{214}$Bi, cosmogenic spallation products such as $^{10}$C, and $2\nu\beta\beta$ itself were found to be the dominant backgrounds for its predecessor KamLAND-Zen~400, which had about half of the enriched Xe mass. While KamLAND-Zen now has the best background index compared to all active and past experiments, it also has the worst detector resolution of experiments mentioned here (Fig.~\ref{fig:res_bkg_exp}). Nevertheless, KamLAND-Zen provides a world-leading $^{136}$Xe limit of $T_{\frac{1}{2}}^{0\nu}$(Table~\ref{tab:ndbd_experiments}). KamLAND-Zen plans to upgrade to KamLAND2-Zen, where 1~ton of enriched Xe will be deployed with a much brighter liquid scintillator and photomultiplier tubes of higher quantum efficiency to improve the energy resolution by a factor of two. KamLAND2-Zen is projected to have a half-life sensitivity of $T_{\frac{1}{2}}^{0\nu}$> $1.1\times10^{27}$~yr in 5 years. 

The Enriched Xenon Observatory 200 (EXO-200)~\cite{EXO-200:2019rkq} was a liquid xenon (LXe), cylindrical time-projection chamber (TPC)  located at the Waste Isolation Pilot Plant (WIPP) in Carlsbad, New Mexico, USA. The liquid-phase TPC provided good energy resolution and low background due to its ability to reconstruct event topology. 

The next-generation Enriched Xenon Observatory (nEXO)~\cite{nEXO:2021ujk}, a successor to EXO-200, will also use an LXe TPC with approximately 5 tons of xenon enriched to 90\% in $^{136}$Xe. nEXO is projected to reach a half-life sensitivity of $T_{\frac{1}{2}}^{0\nu} > 1.35\times10^{28}$~yr with 10 years of data collection. nEXO aims to have an energy resolution of <1\% ($\sigma$) in ROI and plans to reduce the EXO-200 background by a factor of $\sim$1000. The background projections for nEXO are based on its established radioassay data for most component materials and comprehensive particle tracking and event reconstruction simulations. The nEXO collaboration is currently exploring the feasibility of identifying and labelling the daughter atomic element Ba from the double-beta decay of $^{136}$Xe. If this endeavour proves successful, it has the potential to significantly reduce nEXO's background to almost zero in its second phase.

The Neutrino Experiment with a Xenon TPC (NEXT)~\cite{NEXT:2020amj} located at Canfranc Underground Laboratory (LSC), Spain, will use high-pressure xenon-gas time-projection chambers. The experiment aims to capitalise on the naturally low fluctuations in the production of ionisation pairs in xenon gas -- combined with electroluminescence to amplify the ionisation signal -- resulting in an energy resolution of <0.4\% ($\sigma$) at $Q_{\beta\beta}$. Moreover, the tracks left in gaseous xenon have distinct topological features for $0\nu\beta\beta$ events that can be used for background rejection. NEXT-White, a prototype for NEXT-100 and NEXT-1t, has recently successfully demonstrated the TPC technology with a small fiducial volume. NEXT-100 is currently under construction at LSC. NEXT-1t has an estimated half-life sensitivity of >$ 1\times10^{27}$~yr in less than 5~yr with $\sim$1~ton of $^{136}$Xe. NEXT also has its own programme exploring ways to tag the daughter Ba$^{++}$ ions from $^{136}$Xe decay.

%\subsection{LZ/DARWIN}
\subsection{\texorpdfstring{$^{130}$Te ($Q_{\beta\beta}$ = 2527.5 keV)}{}}

The Cryogenic Underground Observatory for Rare Events (CUORE)~\cite{CUORE:2021mvw} is the first ton-scale experiment searching for $0\nu\beta\beta$ decay using low-temperature calorimeters. The detector, located at the LNGS, Italy, consists of an array of 988~$^{nat}$TeO$_2$ crystals; each crystal is equipped with an neutron-transmutation doped germanium (NTD-Ge) thermistor and operated at close to 10~mK. In 2021, CUORE released results corresponding to a ton-year of $^{nat}$TeO$_2$, the largest amount of data ever acquired with a solid-state detector. Low-temperature calorimeters have intrinsically low detector noise, and their detector resolution is comparable to semiconductor detectors. CUORE's background is dominated by the energy-degraded alpha events emanating from detector holders, giving it a high background index. 

The SNO+~\cite{Chen:2021llf} experiment, located at the SNOLAB facility in Canada, will use $^{130}$Te-loaded liquid scintillator. SNO+ developed a novel metal-loading technique using an organic scintillator that keeps the loading stable and enhances the light yield. The initial loading in SNO+ will be
0.5\% $^{nat}$Te by mass, providing a $T_{\frac{1}{2}}^{0\nu} > 2.1\times10^{26}$~yr after 3 years of data taking. The detector-related backgrounds have been measured in two data-taking phases using only water and only liquid scintillator. Deployment of Te-loaded scintillator is planned for 2024. Moreover, recent R\&D efforts have shown that the \isotope{130}{Te} loading can be increased to 3\% by mass with an acceptable scintillator light yield, which would significantly increase SNO+ sensitivity in the future. 

\subsection{\texorpdfstring{$^{100}$Mo ($Q_{\beta\beta}$ = 3034.4 keV)}{}}

The AMoRE~\cite{Kim:2022uce} project aims to search for the neutrinoless double beta decay of $^{100}$Mo using molybdate-based crystals as low-temperature calorimeters. It is located in the Yangyang Underground Laboratory (Y2L) in South Korea. The crystals use metallic magnetic calorimeters (MMCs) to read out phonon signals at milli-Kelvin temperatures. AMoRE-Pilot successfully demonstrated the technology. In 2021, the project moved on to the next phase AMoRE-I, which is currently running with a total of approximately 3~kg of $^{100}$Mo mass. AMoRE-I is housed in the same cryostat used for AMoRE-Pilot. Preliminary results indicate a background rate of $4\times10^{-2}$ counts/kg$\cdot$keV$\cdot$yr in the ROI, and a lower limit of $T_{\frac{1}{2}}^{0\nu} > 1.2\times10^{24}$~yr at 90\% C.L. The next phase, AMoRE-II, is currently under preparation and will operate at Yemilab in South Korea. AMoRE-II will comprise approximately 400 molybdate crystals ($\sim$100 kg of $^{100}$Mo), using both calcium molybdate (CMO) and lithium molybdate (LMO) crystals. The target sensitivity is $T_{\frac{1}{2}}^{0\nu} > 5\times10^{26}$~yr.

CUPID-Mo~\cite{Augier:2022znx} was a demonstrator experiment that employed a dual readout of phonon and scintillating light signals to remove the $\alpha$ background, which is sensitivity-limiting for large-array, low-temperature calorimeter searches like CUORE. This effort followed the success of CUPID-0~\cite{CUPID:2022puj}, the first medium-scale experiment to discriminate $\alpha$ from $\gamma/\beta$ backgrounds with scintillating crystals (Zn$^{82}$Se for CUPID-0). CUPID-Mo took data with 20 LMO crystals flanked by 20 auxiliary low-temperature germanium calorimeters that served as light detectors. Similar to CUORE, all the calorimeters were read out by NTD-Ge thermistors. CUPID-Mo achieved an $\alpha$ rejection efficiency of >99.9\% and an energy resolution similar to the CUORE detectors. The high $Q_{\beta\beta}$, above most environmental $\gamma$ lines, and its $\alpha$ rejection ability enabled CUPID-Mo to establish the feasibility of a larger LMO-based experiment. 

CUORE Upgrade with Particle IDentification (CUPID)~\cite{CUPID:2019imh} is a next-generation, tonne-scale bolometric experiment that will combine the best of the cryogenic infrastructure developed for CUORE and the detector technology developed by CUPID-Mo. A total of 1596 LMO crystals will be installed inside the CUORE cryostat, for a total of 240~kg of $^{100}$Mo. Each crystal will be flanked by two light detectors that will enable $\alpha$ rejection. In the baseline design, the estimated background is <$1\times10^{-4}$ counts/kg$\cdot$keV$\cdot$yr in the ROI, two orders of magnitude lower than CUORE with an energy resolution similar to CUPID-Mo. The projected half-life sensitivity is $T_{\frac{1}{2}}^{0\nu} > 1.4\times10^{27}$~yr at 90\% C.L. for 10 years of livetime. In the future, CUPID aims to push the background index by another factor of 5 with additional purification of the crystals and nearby components, and by reducing the $2\nu\beta\beta$ pile-up background. Eventually, a new cryostat with much more radiopure materials could allow a push to the normal-ordering region. A large-scale infrastructure hosting CUPID-1T would also be uniquely positioned as a multi-isotope observatory, capable of simultaneously deploying multiple cryogenic calorimeters like Zn$^{82}$Se, Li$_2$$^{100}$MoO$_4$,
$^{116}$CdWO$_4$, and $^{130}$TeO$_2$.

\subsection{Nuclear matrix elements (NMEs)}
\label{sec:nme}
Knowledge of the NME for the light-neutrino exchange mechanism is one of the most crucial inputs needed for extracting $m_{\beta\beta}$~\cite{Pompa:2023jxc} from a measured decay rate.  The NMEs are typically calculated using macroscopic many-body nuclear models like the proton-neutron quasiparticle random-phase approximation method, energy-density functional theory, and the interacting boson model or using microscopic models that employ realistic nuclear forces like those from the Nuclear Shell Model or \textit{ab initio} methods. Refs.~\cite{Engel:2016xgb, Ejiri:2019ezh} survey each method with its strengths and weaknesses.  

The microscopic models using \textit{ab initio} methods are computationally complex and have been applied to only light and medium $\beta\beta$ emitters~\cite{Yao:2019rck,Belley:2020ejd}. While macroscopic models have slightly less computational complexity and can cover a wide range of $\beta\beta$ nuclei, they rely on fitting model parameters to a set of experimental observables. The most frequently used nuclear-structure models are the interacting-boson model, the quasiparticle random-phase approximation, energy-density-functional methods, the generator-coordinate method, and the nuclear shell model~\cite{Engel:2016xgb,Agostini:2022zub}. One of the major concerns for nuclear models was the disagreement between the decay rates predicted for $\beta$-decay and $2\nu\beta\beta$-decay, which could be resolved by adjusting $g_{A}$ to a lower value from its nominal value. Since the decay rate for $0\nu\beta\beta$ has a quartic dependence on  $g_{A}$, a reduced value would significantly affect the half-life sensitivity. However, Ref.~\cite{Gysbers:2019uyb} seems to have resolved the discrepancy by including previously neglected nuclear correlations using \textit{ab initio} methods. For many-body theories, calculations continuously improve as our comprehension of various nuclear interactions becomes more refined. Some of these interactions can enhance or suppress the values of the NMEs. For example, Ref.~\cite{Menendez:2011qq} shows that including two-body currents is a key ingredient for calculations and can suppress the $0\nu\beta\beta$ NMEs by $\sim$30\%. On the other hand, recently introduced short-range $0\nu\beta\beta$ NMEs~\cite{Cirigliano:2019vdj} may enhance the rate~\cite{Wirth:2021pij, Jokiniemi:2021qqv} by as much as $\sim$30--50\%.

Given the complexity of the $0\nu\beta\beta$ decay process, one needs independent nuclear wave-function tests to understand and characterise NMEs. There is no one suitable experimental probe that can cover the wide range of momentum components and multipolarities within the nuclear states involved -- the Majorana-neutrino exchange between the two nucleons is localised within $\sim$2~fm, resulting in a momentum spread of 105~MeV/$c$~\cite{Ejiri:2019ezh}. Nevertheless, experimental data from various sources, including studies on ordinary muon capture~\cite{Jokiniemi:2020ydy,Hashim:2021eho}, nucleon transfer reactions~\cite{Freeman:2007mm, Roberts:2013bna}, double gamma decay~\cite{Romeo:2021zrn}, and reactions involving single-charge exchange (SCE) and double-charge exchange (DCE)~\cite{Ejiri:2022ujl}, have been or can be utilised to constrain specific aspects of the calculations related to NMEs in $0\nu\beta\beta$ decay. A good correlation has also been found between the $2\nu\beta\beta$ and $0\nu\beta\beta$ decay NMEs~\cite{Jokiniemi:2022ayc,Horoi:2023uah}. However, in $2\nu\beta\beta$-decay only the low-momentum (few MeV/$c$) components of the nuclear wave functions are probed, and hence, they may not be enough to make deductions for the $0\nu\beta\beta$-decay NMEs. 

Recently, DCE reactions have received significant attention as probes for  $0\nu\beta\beta$-decay NMEs. They share the same initial and final states as $\beta\beta$-decay, can probe a broad range of momentum and multipolarities in intermediate odd-odd isobar nuclei, and are sensitive to nucleon-nucleon interactions, thus resembling some aspects of the $0\nu\beta\beta$-decay mechanism~\cite{NUMEN:2022ton}. The NUMEN (NUclear Matrix Elements for Neutrinoless double beta decay) project aims to systematically investigate various Heavy-Ion-DCE reactions to extract essential information needed for NME calculation~\cite{Cappuzzello:2018wek}. The experimental challenges for NUMEN are immense since the relevant cross sections are tiny (few tens of nano-barns), which requires high ion-beam intensities, with excellent particle identification to select the relevant nuclear channel, and high energy and angular resolution to resolve the transitions to different states from the energy spectra. 
Separate exploratory studies have been performed on different reactions in search of the most promising probe for DCE~\cite{Kisamori:2016jie, Takahisa:2017xry, Matsubara:2013vva}.
Several collaborations also aim to measure ordinary muon capture on double-beta decay-isotopes~\cite{Hashim:2023zwf,Hashim:2021eho,eliza:Monument}.

\section{Additional intersections of neutrino and nuclear physics}
\label{sec:othermeas}

\subsection{Nuclear physics for high-energy neutrino scattering}
\label{sec:high-energy-scattering}

Accelerator neutrino beams with energies of order 0.1--10 GeV are used to explore fundamental mysteries of neutrino physics ranging from the ordering of the neutrino-mass values, the unitarity of the PMNS mixing matrix, and the presence and scale of any CP violation in the neutrino sector. The upcoming, large-scale, long-baseline neutrino experiments Hyper-Kamiokande~\cite{Hyper-Kamiokande:2022smq} (producing neutrinos at J-PARC and observing them at the Kamioka Observatory, Japan) and the Deep Underground Neutrino Experiment (DUNE, producing neutrinos at Fermilab and observing them at SURF, USA)~\cite{DUNE:2022aul} rely on such beams. Nuclear physics enters these measurements both via the neutrino source, typically a decay-in-flight pion beam produced by protons striking a thin target, and via the observation of a neutrino interaction in a detector medium. Conventional neutrino beams are not mono\-energetic, although the detector location can be chosen to sample a particular region of the energy distribution. The neutrino energy $E_\nu$ must be reconstructed from the event data and is thus affected not only by the energy-dependent cross section but also by uncertainties in the modelling of the target nucleus, including resonant processes; hadronic final-state interactions within the nuclear medium; secondary interactions outside the nucleus; and the final-state topology~\cite{NuSTEC:2017hzk}. 

Nuclear processes are now recognised as a major limiting factor on the sensitivity of large neutrino-oscillation experiments, e.g.~\cite{Benhar:2015wva,Balantekin:2022jrq}. In principle, they are included in the Monte-Carlo event generators that simulate neutrino-nucleus interactions~\cite{Campbell:2022qmc}. 
However, the accuracy of these simulations is hindered by a lack of experimental data and precise theoretical calculations, driving substantial investment in both dedicated experiments and sophisticated near-the-source detectors that simultaneously normalise the flux for long-baseline experiments and pursue independent physics measurements. Indeed, standard event generators are discrepant with measured neutrino interactions, e.g.~\cite{MINERvA:2019kfr,NOvA:2020rbg}. These discrepancies may point to insufficient constraints on the axial coupling to the nucleon as well as to nuclear effects~\cite{Nikolakopoulos:2022tut}. 
Future neutrino-scattering measurements -- whether via the near detectors of short- or long-baseline oscillation experiments, or via dedicated experiments such as ANNIE~\cite{ANNIE:2019azm}, NINJA~\cite{NINJA:2020gbg}, or nuSTORM~\cite{nuSTORM:2022div} -- will help illuminate these issues, although in these measurements it is challenging to disentangle the specific physics mechanisms underlying any discrepancies. Neutrino-nucleon scattering measurements on hydrogen or deuterium targets may help disambiguate results~\cite{Alvarez-Ruso:2022ctb}. 

Pion-nucleus scattering measurements probe hadronic final-state interactions within the target nucleus. Recent and future experimental efforts in this line focus on the same targets used by current and next-generation neutrino detectors: water~\cite{Yamauchi:2021cet}; carbon~\cite{DUET:2015ybm, DUET:2016yrf}; and argon~\cite{LArIAT:2019kzd, DUNE:2021hwx}. World $\pi^\pm$ data is already being used to tune the NEUT intranuclear cascade model~\cite{PinzonGuerra:2018rju}. 

Electron-nucleus scattering measurements exploit both high statistics and precise control of incident electron energies and final-state kinematics to probe the vector part of lepton-nucleus interactions, as reviewed in Ref.~\cite{Ankowski:2022thw}. A recent test of neutrino energy-reconstruction techniques against electron scattering data on ${}^4$He, ${}^{12}$C and ${}^{56}$Fe revealed significant discrepancies~\cite{CLAS:2021neh}. Current~\cite{JeffersonLabHallA:2022cit,JeffersonLabHallA:2022ljj} and planned~\cite{Ankowski:2019mfd} measurements explore nuclear spectral functions and lepton-nucleus cross sections. Extensions of neutrino event-generators to predict electron-scattering observables are underway, \textit{e.g.}\,Ref.~\cite{electronsforneutrinos:2020tbf}.

In addition to constraining the nuclear physics of neutrino interactions, neutrino-scattering experiments also probe core questions in nuclear physics. For example, MINERvA recently made the first direct measurement of the free-proton axial-vector form factor, based on an analysis of $\bar{\nu}_\mu + p \rightarrow \mu^+ + n$ events~\cite{MINERvA:2023avz}, and the planned DUNE near detector will measure the electroweak mixing angle $\sin^2\theta_W$ and probe isospin physics in hydrocarbon and argon targets~\cite{DUNE:2021tad}.

\subsection{Nuclear physics from low-energy neutrino scattering}
\label{sec:low-energy-scattering}

At low energies, neutrino scattering becomes a probe of nuclear structure. Coherent elastic neutrino-nucleus scattering (CEvNS), a neutral-current interaction with a relatively large cross section, probes the neutron distribution within a nucleus~\cite{Abdullah:2022zue}. The complete COHERENT data set on CsI~\cite{COHERENT:2021xmm} has been used to determine the averaged neutron radius $R_n$ for Cs and I to within about 6\%; the precision can be improved by combination with atomic parity-violation data~\cite{AtzoriCorona:2023ktl}. Appropriate nuclear targets for CEvNS allow low detection thresholds; by contrast, nuclear targets for $R_n$ via parity-violating electron scattering (${}^{27}$Al~\cite{Qweak:2021ijt}, ${}^{48}$Ca~\cite{CREX:2022kgg}, ${}^{208}$Pb~\cite{PREX:2021umo}) are chosen for high-lying nuclear excited states and for robustness under intense irradiation. Both neutrino- and electron-scattering techniques avoid the model dependencies of hadronic probes~\cite{Thiel:2019tkm} while illuminating complementary regions of the neutron-distribution landscape. Combined with measurements of the proton radius, these results explore the nuclear symmetry energy and inform our understanding of neutron stars~\cite{Horowitz:1999fk}. 

Neutrinos with energy of order 10~MeV are an important driver of supernova nucleosynthesis, interacting with abundant nuclei in the collapsing star to produce rare, often neutron-poor isotopes~\cite{Woosley:1990, Sieverding:2019qet}. Direct measurements of these charged-current interactions are useful inputs to models of this $\nu$-process nucleosynthesis. Supernova neutrinos of $\mathcal{O}(1-10)$~MeV will appear in worldwide detectors through a variety of detection channels~\cite{Scholberg:2012id}, many of which will benefit from dedicated measurements to reduce uncertainties on supernova dynamics and other observables. Two recent charged-current measurements from COHERENT --  ${}^{nat}\mathrm{Pb}(\nu_e, \mathrm{X}n)$~\cite{COHERENT:2022eoh} and ${}^{127}\mathrm{I}(\nu_e, \mathrm{X}n)$~\cite{COHERENT:2023ffx} -- show significant deficits relative to theoretical predictions in the MARLEY framework~\cite{Gardiner:2021qfr}, highlighting the need for further work. 

\subsection{Reactor antineutrinos and nuclear physics}
\label{sec:reactornu}
%\todo{0.5 pg -- target hit!}

Nuclear reactors produce copious amounts of $\bar{\nu}_e$ via beta-decay chains fed by fission reactions in the core. The first experimental discovery of neutrinos was made at the Savannah River reactor~\cite{Cowan:1956rrn}. Since then, reactor antineutrinos have been instrumental in completing the picture of three-neutrino oscillation. The large-scale Jiangmen Underground Neutrino Observatory (JUNO), under construction in China, will observe reactor antineutrinos~\cite{JUNO:2021vlw}.

In the last decade, high-precision reactor experiments independently observed two anomalies sometimes taken as evidence of sterile neutrinos: a $\sim 5-6\%$ flux deficit relative to the Huber-Muller prediction~\cite{Huber:2011wv, Mueller:2011nm} based on the conversion of summed beta spectra to antineutrino spectra, and an excess of antineutrinos at about 5~MeV~\cite{RENO:2016ujo, DayaBay:2019yxq, DoubleChooz:2019qbj}. Extensive experimental and theoretical work, including new beta-spectrum measurements~\cite{Kopeikin:2021ugh}, a reconsideration of decay-heat measurements~\cite{Sonzogni:2023xjh}, and studies of the neutrino flux for different fuel compositions~\cite{DayaBay:2017jkb, RENO:2018pwo, Stereo:2021wfd}, suggest attribution of the flux deficit to biases in the model inputs. Meanwhile, investigations of the 5-MeV excess revealed errors in nuclear databases~\cite{Sonzogni:2016yac}; the precise origin of this feature remains unclear, but the likely presence of contributions from all primary fission isotopes suggests a common error in the flux prediction~\cite{PROSPECT:2022wlf}. Precise reactor-antineutrino measurements are improving our understanding of nuclear fission.

The impossibility of shielding antineutrinos gives them an appealing possible application in nuclear non-proliferation, recently reviewed in Ref.~\cite{Bernstein:2019hix}: in principle, measuring characteristic antineutrino spectra allows the detection of a covert fission plant, or non-invasive monitoring of spent fuel or reactor operations. However, neutrino detection (especially in a high-background reactor environment) requires both significant financial investment and exposure time, and is likely impractical without facility cooperation. The Nu~Tools study, based on discussions with end users in nuclear energy and nuclear security, found that neutrino monitoring would most likely be useful in the context of future nuclear deals; assay of spent fuel in dry casks; and future advanced reactors where traditional accountancy methods cannot be used~\cite{Akindele:2021sbh}. Further development is needed for practical neutrino monitoring.

\subsection{Searching for sterile neutrinos with nuclear physics}
\label{sec:sterilenu}

\begin{figure}[tbp]
    \centering
    \includegraphics[width=0.6\textwidth]{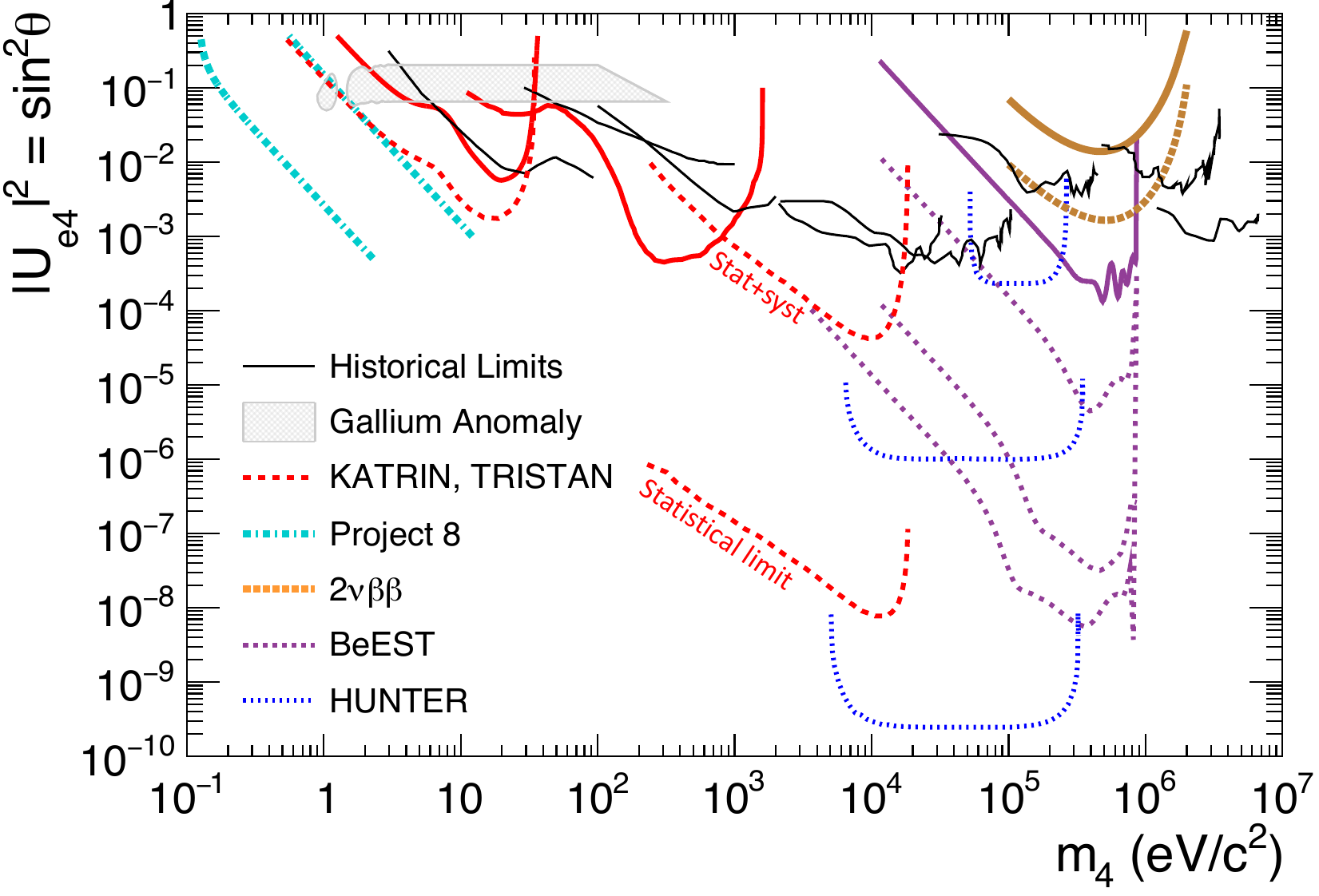}
    \caption{Achieved (solid) and projected (dotted) exclusion curves for sterile neutrinos from $\beta$-decay experiments, along with the parameter space preferred by the gallium anomaly ($2\sigma$ contours). Adapted from Ref.~\cite{Acero:2022wqg}.} %CC-BY license, https://arxiv.org/abs/2203.07323
    \label{fig:betadecay-sterile-excl}
\end{figure}

Apart from reactor-based searches (Sec.~\ref{sec:reactornu}), nuclear physics is key to non-oscillation-based searches for sterile neutrinos.
The spectrum from a beta or electron-capture decay (Sec.~\ref{sec:numass}) is, in principle, a superposition of spectra: one for each neutrino-mass state, the mass value of which shifts the endpoint of the spectrum. Although the splittings of the three known mass states are too small for current-generation measurements to resolve, the presence of a fourth, widely separated neutrino-mass value $m_4$ will generate a kink-like spectral distortion at $E_0 - m_4$, where $E_0$ is the spectral endpoint. 
KATRIN has searched for this sterile-neutrino signature at both the eV scale~\cite{KATRIN:2022ith}, excluding significant portions of the parameter space that could explain the reactor flux deficit, and (using a commissioning data set) at the keV scale~\cite{KATRIN:2022spi}. A planned future phase of KATRIN will perform higher-sensitivity searches for keV-scale sterile neutrinos with high-rate, deep spectral measurements enabled by the TRISTAN detector upgrade~\cite{KATRIN:2022ayy}. Planned Project~8 operations would further improve sensitivity at the eV scale~\cite{Project8:2022wqh}.

Neutrino-mass experiments favour low $E_0$, setting a ceiling on any observable value of $m_4$. Two experiments aim to push past that ceiling by precisely measuring the kinematics of electron-capture decays with higher $E_0$: ${}^7$Be (BeEST, first limit set in Ref.~\cite{Friedrich:2020nze}) and ${}^{131}$Cs (HUNTER, planned~\cite{Martoff:2021vxp}). This allows kinematic reconstruction of the neutrino four-momentum and corresponding sensitivity to a heavy mass state. The presence of a sterile neutrino would also affect the electron kinematics in $2\nu\beta\beta$ decay; Bolton et al.~\cite{Bolton:2020ncv} explore the corresponding sensitivity of $0\nu\beta\beta$-decay searches (Sec.~\ref{sec:neutrinoless}) to sterile neutrinos. 
Fig.~\ref{fig:betadecay-sterile-excl} shows the existing (solid) and projected (dotted) limits on sterile-neutrino mixing $|U_{e4}|^2$ and mass $m_4$ from beta-decay and double-beta-decay experiments. %energy reconstruction only for 2vbb

Beyond beta decays, nuclear physics may also be central to the longstanding gallium anomaly. When a high-intensity $\nu_e$ source irradiates a gallium target, ${}^{71}\mathrm{Ga}(\nu_e,e^-){}^{71}\mathrm{Ge}$ interactions may be counted using radiochemical methods. The combined result of historical (GALLEX~\cite{GALLEX:1997lja, Kaether:2010ag} and SAGE~\cite{SAGE:1998fvr, Abdurashitov:2005tb}) and modern (BEST~\cite{Barinov:2022wfh}) experiments is a significant deficit in the observed $\nu_e$ rate. An overestimation of the nuclear-interaction cross section has been proposed as an alternative explanation to the oscillation of $\nu_e$ into a sterile flavour; however, a recent re-calculation of corrections to the cross section shows only modest effects, and the well-measured ground-state transition prohibits large changes~\cite{Elliott:2023xkb}.
 Followup experiments with intense $\nu_e$ and $\bar{\nu}_e$ sources, or the realisation of a fundamental problem with the nuclear-interaction calculation, could help resolve this anomaly.

\subsection{Synergy between neutrino-nuclear physics and quantum sensing}
\label{sec:quantum_synergy}

%\st{Quantum sensors are gaining prominence in cutting-edge neutrino-nuclear experiments}. 
Quantum sensing, broadly used to describe the use of quantum objects or phenomena for measuring physical quantities, whether classical or quantum, is playing a crucial role in advancing precision measurements and gaining prominence in cutting-edge neutrino-nuclear experiments. Transition-edge sensors (TES) are used to achieve precise energy spectra from nuclear-$\beta$ decay (as seen in HOLMES~\cite{HOLMES:2016spk}); explore coherent neutrino nuclear scattering (NUCLEUS~\cite{NUCLEUS:2022zti} and RICOCHET~\cite{Ricochet:2023yek}); and pursue next-generation $0\nu\beta\beta$ searches like CUPID~\cite{Singh:2022rck}. Simultaneously, superconducting tunnel junctions (STJs) are instrumental in BeEST's~\cite{Friedrich:2020nze} investigation of phenomena such as sterile-neutrino states, while ECHo~\cite{Velte:2019jvx} and AMoRE~\cite{Kim:2022uce}  are developing arrays of metallic magnetic calorimeters (MMCs) for a neutrino-mass measurement and a $0\nu\beta\beta$-decay search, respectively. Project~8~\cite{Project8:2022hun} aims to develop superconducting parametric amplifiers near the quantum limit for a tritium-based $m_\beta$ measurement, and the QTNM project~\cite{qtnmkit2023} aims to push still further by adding quantum-sensor magnetometry. Experimental needs demand faster sensor response times, expanded channel capacity, and more efficient multiplexing capabilities to enable the simultaneous readout of multiple sensors on a single line. Cryogenic hardware employed for quantum-sensor readout predominantly relies on superconducting microwave resonators and superconducting quantum interference devices (SQUIDs). These readout technologies have the potential for broader applications beyond quantum sensing, such as interfaces with large qubit arrays.

On the flip side, the development of low-radioactivity techniques, primarily designed to explore rare phenomena like $0\nu\beta\beta$ decays and dark-matter searches, offers a distinctive opportunity to address the impact of ionising radiation on quantum sensors for quantum computers that significantly rely on the quantum phenomena of coherence and entanglement. Recent works have explored low-radiation materials, shielding, and underground quantum-circuit locations to reduce the effect of ionising radiation on superconducting qubits~\cite{Vepsalainen:2020trd,Cardani:2020vvp}. Ionising radiation can lead to correlated errors, which pose a significant challenge for error correction and jeopardise the performance of quantum algorithms~\cite{Wilen:2020lgg, McEwen:2021wdg,martinis2021saving}. 
Hence, understanding the physics of how ionising radiation thermalises in qubit devices is crucial for successful mitigation and for advancing quantum error correction at scale. While operating quantum computers deep underground, with extensive shielding material around them, may not be feasible for large-scale applications, deep underground nuclear-physics facilities present a unique opportunity to research radiation effects within a controlled environment. 

\section{Conclusion}
\label{sec:conclusion}
Since 1930, when Pauli postulated the neutrino's existence to restore energy-momentum conservation in nuclear beta decays, we have learned a great deal about this ghostly particle.
The 1956 discovery of neutrinos via inverse beta decay at a nuclear fission plant was a triumph of experimental neutrino physics.  However, fundamental questions about neutrinos remain open despite significant theoretical and experimental progress. In this paper, we have reviewed the many connections between nuclear physics and neutrino physics, which illuminate questions in both areas.  

In the Standard Model, the neutrinos are the only massless fermions.  We now know that neutrinos in fact have mass, and we have seen how current and future beta-decay and electron-capture experiments can probe the neutrino mass scale below the current limit via direct kinematic measurements in various nuclei.

Although we have observed the second-order $2\nu\beta\beta$ decay in certain nuclei, we have yet to detect lepton-number-violating $0\nu\beta\beta$ decay in any nucleus.  We have discussed future experiments with ever larger sizes and higher sensitivities that will investigate the possibility that neutrinos are their own antiparticles (i.e., whether they are Majorana fermions).

Finally, we discuss additional ways in which nuclear physics and neutrino physics intertwine, from final states in high-energy interactions, to nuclear structure, to searches for sterile neutrinos, to cutting-edge developments in quantum sensing.
The quest to understand neutrino properties is a multi-disciplinary effort, and the nucleus is a critical laboratory for many of these endeavours.

\vskip6pt

\ack{DSP acknowledges support from the U.S. Department of Energy (DOE), Office of Science, under Award Numbers DE-SC0010118, DE-SC0019304 and DE-SC0022125. AP is supported by the U.S. DOE under Federal Prime Agreement DE-AC02-05CH11231. VS is supported by the U.S. DOE, Office of Science, under grant DE-FG02-00ER41138. We thank Alexey Lokhov, Moritz Machatschek, Lisa Schl\"{u}ter, Pranava Teja Surukuchi, and Kathrin Valerius for their useful contributions and suggestions.}

%%%%%%%%%% Insert bibliography here %%%%%%%%%%%%%%

\bibliographystyle{RS}
\bibliography{main}

\end{document}